\documentclass[preprint,preprintnumbers,nofootinbib,aps,amsfonts]{revtex4}
\usepackage{verbatim}
\usepackage[hyperfootnotes=false]{hyperref}
\begin{document}
\title{A Simple Bound on the Error of Perturbation Theory in Quantum Mechanics}
\author{Daniel Harlow}
\email {dharlow@stanford.edu}
\affiliation{Department of Physics, Stanford University, Stanford, CA, 94305}
\preprint{SU-ITP-09/20}
\begin{abstract}
I provide a straightforward proof that a simple harmonic oscillator perturbed by an (almost) arbitrary positive interaction has a perturbative expansion for any finite-time Euclidian transition amplitude which obeys the following result: the difference of the sum of the first N terms of the series and the exact result is bounded in absolute value by the next term in the series.  The proof makes no assumptions about either the strength of the interactions or the convergence of the perturbation series.  I then argue that the result generalizes immediately to a much broader class of quantum mechanical systems, including bare perturbation theory in quantum field theory.  The case of renormalized perturbation theory is more subtle and remains open, as does the generalization to energy levels and connected S-matrix elements.   
\end{abstract}
\maketitle

\section{Introduction}
Perturbation theory has historically been the most useful tool we possess in analyzing quantum mechanical systems for which we have no exact solution.  In particular in elementary particle physics its predictions have been confirmed to astonishing accuracy, for example in the four-loop calculations of the anomolous magnetic moment of the electron and muon.  However from a mathematical point of view it is clear that for many interesting systems the perturbation series is a tool of fundamentally limited value.  For models as simple as the quartic anharmonic oscillator, the series is known to diverge for all non-zero values of the coupling, and moreover well-known ``non-perturbative'' effects such as instantons and solitons depend on the coupling constant in ways that are invisible to perturbation theory.  These issues were not understood until the 1970's, after which consensus developed around a standard folklore which I will review briefly in the following section.  I expect that this folklore is basically correct, but there is a key logical step which I was unable to find in the literature: to really evaluate the usefulness of a given truncated perturbative expansion, we need to know how far away we are from the exact result.\footnote{In several ancient papers \cite{kato1,kato2,Simon:1970mc} it is intricately proven using functional analysis that for some fairly general systems the perturbation series for the energy levels and eigenstates is \textit{asymptotic} in the sense that the error of truncating the series at fixed $N$ vanishes like $g^{N+1}$ for sufficiently small $g$.  It would however be preferable to have an explicit and calculable bound at finite $g$, especially given that for many experiments the perturbation is not something we can tune at will.}
Indeed without a bound of this sort, perturbation theory is an uncontrolled approximation and its use is logically questionable, especially in light of the mathematical difficulties I mentioned above.  We may of course provisionally use its remarkable experimental success to justify it, but it would be more intellectually satisfying to find a mathematical explanation.  

\section{The Standard Lore}
The basic assumption of what could be called ``naive perturbation theory'' is that the transition amplitudes of a quantum mechanical system with Hamiltonian $H_0+gV$ can be expanded in a convergent power series in the coupling constant $g$.  Moreover it is not hard to prove that a power series in $g$ that converges to the amplitude for $g\leq\epsilon$, with $\epsilon$ some positive real number, also defines a holomorphic function in the disk $|g|<\epsilon$.  Thus we are really assuming that the transition amplitudes can be continued to holomorphic functions of $g$ about the origin.  It is only in such circumstances that the perturbation series can converge, and only given such convergence that we can make the error as small as we like by adding more and more terms.   

In fact for the anharmonic oscillator we can see via two separate arguments that the perturbation theory must diverge \cite{Simon:1970mc,Bender:1969si,Bender:1973rz}.  The propagator for this theory has path integral representation:\footnote{In this paper I will always assume that the path integral is explicitly discretized and is thus just a finite product of ordinary integrals.  However since the bound that I will derive will be true for all finite values of the regulator, it will be true for the limiting case as long as the limit exists.  In ordinary quantum mechanics we know that it almost always does, but in quantum field theory I will opt to keep the regulator explicit (at least in mind if not in notation).}
\begin{equation} 
D(g,x_i,x_f,T)\equiv\langle x_f|e^{-TH}|x_i \rangle = \int\!\left[ \mathcal{D}x\right]_{x_i}^{x_f} \,e^{-\int_0^T\!dt\,\left(\frac{\dot{x}^2+x^2}{2}+gx^4\right)} 
\label{oscillator}
\end{equation}
The perturbation series is defined by expanding the exponential of the interaction and cavalierly exchanging the order of summation and integration:
$$D(g,x_i,x_f,T)\text{``=''}\sum_{k=0}^\infty D_k g^k$$
\begin{equation}
D_k\equiv \frac{(-1)^k}{k!}\int\!\left[ \mathcal{D}x\right]_{x_i}^{x_f} \,\left(\int_0^T\!dt\,x^4\right)^k e^{-\int_0^T\!dt\,\left(\frac{\dot{x}^2+x^2}{2}\right)}
\end{equation}
Since this section concerns standard knowledge I will simplify the notation by sketching both non-convergence arguments for a single integral whose behaviour turns out to be analogous to that of the path integral.\footnote{The steepest descent arguments for the path integral case require a discussion of instanton solutions of the equations of motion.  The reader who wants more details can consult \cite{Itzykson:1980rh,ZinnJustin:2002ru,Brezin:1976vw}.} Define:
\begin{equation}
A(g)\equiv\int_{-\infty}^{\infty}\!dx\,e^{-\left(\frac{x^2}{2}+\frac{g}{4}x^4\right)}
\end{equation}
 
The first argument begins by showing that $A(g)$ is analytic in the $g$ plane for $Re[g]>0$.  For $Re[g]<0$ the integral doesn't exist, but we can prove that it can be analytically continued by smoothly rotating the contour of integration opposite to the phase of $g$ to ensure that $gx^4>0$.  Finally we show that this analytic continuation has a branch cut on the negative real axis, whose discontinuity can be estimated by the method of steepest descent to be proportional to $e^{-\frac{1}{4|g|}}$.  Since the function is thus not holomorphic at the origin, it cannot have a convergent power series. 

The second argument is to estimate the large orders of perturbation theory and show explicitly that the series diverges.  We thus need to define the perturbative expansion:
$$A(g)\text{``=''}\sum_{k=0}^\infty A_k g^k$$
\begin{equation}
A_k\equiv \frac{(-1)^k}{4^kk!}\int_{-\infty}^{\infty}\!dx\,x^{4k} e^{-\frac{x^2}{2}}
\end{equation}
This case is so simple that each order can be evaluated explicitly in terms of gamma functions, but a method which can be generalized to path integrals is to use the method of steepest descent to approximate $A_k$ in the limit of large k.  This reveals that $A_k\sim k!$ for large k, with subleading factors that match the result of applying Stirling's approximation to the exact gamma function expression.  Thus $A_k g^k$ will grow arbitrarily large with $k$, regardless of the smallness of $g$.

In fact we can also show that the $k$ for which $A_k g^k$ is minimized also has $A_k g^k \sim e^{-\frac{1}{4g}}$.  This result, along with the divergence of the series, is often used to justify claims that the perturbation series is asymptotic and that the size of the smallest term is the size of ``non-perturbative'' effects.  However to really prove these results we need to bound the error of summing a finite number of terms of the perturbation series.  A bound of this sort for the simple integral is presented but not derived in \cite{ZinnJustin:1980uk}.  I will now provide such a bound for the path integral.  
\section{Proof of the theorem}
We begin with the exact path integral expression from equation (\ref{oscillator}), but generalizing the interaction to an arbitrary positive function $f(x)$,\footnote{$f(x)$ is also restricted by the requirement that functional integrals in the perturbation expansion exist.  This will be taken into account in the statement of the theorem below} defining $I[x]\equiv \int_0^T\!dt\,f(x)$, and suppressing unnecessary notation we have:
\begin{equation}
D(g)=\int\!\left[ \mathcal{D}x\right]\,e^{-\left(S_0+gI\right)}
\end{equation}
The error of $N$ orders of perturbation theory is then given by:
\begin{equation}
D(g)-\sum_{k=0}^N D_k g^k=\int\!\left[ \mathcal{D}x\right]\,e^{-S_0} \sum_{k=N+1}^{\infty} \frac{(-gI)^k}{k!}
\end{equation}
Here expanding the exponential inside the path integral is perfectly justified, as is moving the finite sum of the first N terms through the path integral.  Pulling out a common factor and relabeling the sum, we get:
\begin{equation}
D(g)-\sum_{k=0}^N D_k g^k=(-g)^{N+1}\int\!\left[ \mathcal{D}x\right]\,I^{N+1} e^{-S_0} \left[ \sum_{k=0}^{\infty} \frac{1}{(k+N+1)!}\left(-gI\right)^k\right]
\label{error}
\end{equation}
Now the key step: the sum in brackets has an integral representation:
\begin{equation}
\sum_{k=0}^{\infty} \frac{1}{(k+N+1)!}\left(-x\right)^k=\frac{1}{N!}e^{-x}\int_0^1\!dt\,t^Ne^{xt}
\end{equation}
The reader can verify this by putting both sides into \textit{Mathematica}, but I will also sketch a proof in the appendix.  Inspecting the right hand side of this identity we see that for $x\geq0$ the integrand is positive, increases monotonically throughout the range of integration, and is bounded by the function $t^Ne^x$.  As such we can easily prove that:
\begin{equation}
0\leq \sum_{k=0}^{\infty} \frac{1}{(k+N+1)!}\left(-x\right)^k\leq \frac{1}{(N+1)!}
\end{equation}
Now going back to our expression for the error in equation (\ref{error}), we see that the integrand of the path integral is positive, and given our assumption of positive $f(x)$ it can be bounded from above by replacing the sum in the brackets by $\frac{1}{(N+1)!}$.  We have thus proven:
\begin{equation}
|D(g)-\sum_{k=0}^N D_k g^k|\leq \frac{g^{N+1}}{(N+1)!}\int\!\left[ \mathcal{D}x\right]\,I^{N+1} e^{-S_0}=|D_{N+1}g^{N+1}|
\end{equation}

This ends the proof for the simple oscillator system, but the argument depended very little on the forms of $f(x)$ and $S_0$.  For example if we had considered a system of many oscillators coupled by a positive interaction, the steps would have been identical.  In fact the only requirement on $S_0$ is that it falls off fast enough for the integrals to exist.  We have thus really proven:

THEOREM: Given two functionals $S_0[x_i]$, $I[x_i]$ of a finite set of functions such that $I$ and $S_0$ are well-defined on all functions $x_i(t)$, that $I\geq 0$, and such that the following functional integrals exist:\footnote{Here recall from footnote 2 that I mean they exist after being discretized.  This theorem will be agnostic about the limit as the regulator is removed, but for Gaussian $S_0$ the limit can usually be taken and the theorem will be then true for the limits.} $$Z(g)=\int\!\left[ \mathcal{D}x_i\right]\,e^{-\left(S_0[x_i]+gI[x_i]\right)}$$ 
$$Z_k=\frac{(-1)^k}{k!}\int\!\left[ \mathcal{D}x_i\right]\,I[x_i]^k e^{-S_0[x_i]}\qquad0\leq k \leq N+1$$
Then:
\begin{equation}
|Z(g)-\sum_{k=0}^N Z_k g^k|\leq|Z_{N+1}g^{N+1}|
\label{bound}
\end{equation}
 
\section{Application to Field Theory}
The theorem formulated in the last section is general enough to include the case of regulated quantum field theory.  As an example consider a self-interacting scalar field theory in $d\leq 4$ dimensions defined on a lattice in a box: the functions being integrated over are $\phi_{\vec{x}}(t)$, the boundary conditions are chosen for example to be periodic, and the functionals are:
$$S_0=\int_0^T\!dt\,\sum_{\vec{x}\in \text{lat.}}\frac{1}{2}\left(Z_0(\partial \phi)^2+m_0 \phi^2 \right) $$
$$I=\int_0^T\!dt\,\sum_{\vec{x}\in \text{lat.}}\phi^4$$
The functional integrals $Z$ and $Z_k$ exist, and thus the bound (\ref{bound}) applies.

Unfortunately, the perturbation series whose coefficients are $Z_k$ is not the perturbation series we are actually interested in.  As we remove the lattice and box regulators it will almost certainly diverge, and even if we do not the strong dependence of the higher orders on extremely high energy physics violates our sensibility.  We must instead introduce renormalized perturbation theory before we can produce physically meaningful results.  However the renormalized coupling enters the functional integral in a subtle way: we demand that the bare couplings $Z_0$\footnote{The field renormalization coefficient, not the 0th order of perturbation theory!}, $m_0$, and $g_0$ depend on both the regulator and the renormalized coupling and mass in such a manner as to satisfy Wilson's renormalization group equations \cite{Wilson:1973jj}.  This functional dependence can be extremely complex, and there is absolutely no guarantee that the simple manipulations of the previous section will generalize.  Moreover for renormalized perturbation theory even the standard large-order arguments I sketched in section II run into difficulties  related to ``renormalon'' singularities that can dominate the instanton contributions to large orders \cite{Weinberg2, ZinnJustin:2002ru}.

Nonetheless I expect that the situation is better than it appears.  The problem with our calculation is that we are attempting to use a single renormalization scheme to produce a perturbation series for all physical processes.  The lesson of the renormalization group is exactly that this is how we should NOT proceed.  Instead we should consider the functional integral expression for a specific ``infrared-safe'' observable at a specific energy scale, and then choose a renormalization scheme that is appropriate for that scale.  It is only this perturbative expansion that really must obey the error bound, and it seems plausible to me that it will.  I believe that by carefully examining this perturbation series and its relation to the bare perturbation series, it should be possible to generalize my argument above to the renormalized series, but I have not seen a simple way to do so.
  
\section{Energy Levels and Connected S-matrix Elements}

So far I have discussed Euclidian transition amplitudes at finite time, but in fact these are not the objects that are normally studied using perturbation theory: indeed they have a ``large time'' infrared issue as I will now argue.\footnote{This problem was emphasized to me by E. Witten.}  For most experiments the relevant time interval in the transition amplitude is much larger than the microphysical timescale of the system in question.  Moreover this time-dependence becomes stronger and stronger at higher orders of perturbation theory, essentially because at higher orders we can have more and more disconnected Feynman diagrams and each diagram will be proportional to a power of $T$ with the power given by the number of connected components of the diagram.  This means that regardless of the smallness of $g$, for large $T$ the perturbation series will be useless.  This is consistent with the theorem proven above since the bound becomes trivial in the large $T$ limit, but it also means that the theorem is not quite what we would want.  Thus even in the case of simple quantum mechanics we need to do some sort of infrared resummation to get a useful result.  

The observables that are usually approximated by perturbation series are energy levels and connected S-matrix elements, so it is for these that we would really like to find a non-trivial bound.  A colleague and I have spent some time thinking about this, \cite{DongHarlow} but so far we have no decisive result.  I will say a little about how the problem can be approached, and quote an intermediate result we have obtained.  

The infrared resummation is easy to perform in these cases; it amounts to simply taking the logarithm of the perturbation series.  For example if $Z=tr\left[e^{-TH}\right]$ and $Z_0=tr\left[e^{-TH_0}\right]$, then to find the shift in the ground state energy (assuming a discrete spectrum in the vicinity of the ground state) we have:

\begin{equation}
E-E_0=-\lim_{T\to\infty}\frac{1}{T}\log\frac{Z}{Z_0}
\label{energyshift}
\end{equation}

A similar result holds for the generating functional of connected correlation functions.\footnote{See for example chapter 4 of \cite{Peskin:1995ev}.  Correlation functions are then related to the S-matrix via the usual LSZ argument.}  We would thus like to find a way to bound the error of the perturbative expansion for the \textit{logarithm} of a path integral.  The result that we might expect to be true is that the previous result continues to hold but with with only connected diagrams appearing on the right-hand side of equation \ref{bound}.  Indeed using methods similar to those of section three (and under the same assumptions) we have shown that: \cite{DongHarlow}
\begin{equation}
|\log Z(g)-\log Z_0|\leq g\left|\frac{Z_1}{Z_0}\right|
\end{equation}
This result is quite interesting, for example if we combine it with equation \ref{energyshift} we see that the absolute value of the ground state energy shift is bounded by the absolute value of the first order correction.\footnote{The reader might be surprised by this since there are well-known examples where the first correction to the ground state energy vanishes, but these always involve either fermions or an interaction unbounded from below, so the assumptions of the theorem are violated.  In fact this bound for the ground state energy also follows from a simple variational argument, as was pointed out to me by L. Susskind.}  The reader can see that the fraction on the right-hand side cancels all disconnected diagrams to first order in perturbation theory, so the large $T$ behaviour of both sides is linear and the bound is now nontrivial in the large $T$ limit.  Unfortunately with no other restrictions this bound does not generalize to higher orders; in fact my colleague and I have found (admittedly unphysical) counterexamples.  Nonetheless it does seem to be obeyed in a variety of interesting examples, so we expect that with a sufficient restriction it may still be true.

\section{Conclusion}

The situation is thus mixed.  For short-time transition amplitudes in simple quantum mechanical systems we have a very satisfying result; in particular it assures us that as long as we find successive terms in perturbation theory to be smaller and smaller, we are getting closer and closer to the exact answer and should keep computing higher orders.  Given the potentially divergent nature of the series, this was not at all obvious beforehand.  Moreover we see that the smaller the coupling, the more accuracy we can achieve before the series begins to diverge.  This justifies assigning the label of ``size of non-perturbative effects'' to the smallest term in the series.  In fact the theorem has as an obvious corollary the usual statement that the perturbation series is \textit{asymptotic} in the mathematical sense, so in this case we have managed to derive both parts of the lore stated at the end of section II.  However so far there is no satisfactory generalization to energy levels and connected S-matrices, as well as no obvious way to take into account renormalization.  For the first of these I have quoted a promising initial result, but it is fair to say that both questions remains open.  
\acknowledgments
I would like to thank Tom Banks, Bernd Jantzen, Daniel Green, Mahdiyar Noorbala, and especially Xi Dong for helpful comments and discussion, and I would also like to thank Edward Witten for very useful correspondence on a previous version of this work, as well as Lenny Susskind for both discussion and encouragement to publish this result.  I am supported by a Melvin and Joan Lane Stanford Graduate Fellowship.
\appendix
\section{Proof of the Identity}
In the text we used the following identity:
\begin{equation}
\sum_{k=0}^{\infty} \frac{1}{(k+N+1)!}\left(-x\right)^k=\frac{1}{N!}e^{-x}\int_0^1\!dt\,t^Ne^{xt}
\label{identity}
\end{equation}
To derive it, observe:\footnote{A shorter alternative proof using facts about (Euler) beta functions was suggested to me by B. Jantzen.}
\begin{eqnarray*}
N!\,e^x\sum_{k=0}^{\infty} \frac{1}{(k+N+1)!}\left(-x\right)^k&=&N!\frac{e^x}{(-x)^{N+1}}\sum_{k=N+1}^\infty \frac{(-x)^k}{k!}\\
&=&N!\frac{e^x}{(-x)^{N+1}}\left(e^{-x}-\sum_{k=0}^N\frac{(-x)^k}{k!}\right)\\
&=&\frac{N!}{(-x)^{N+1}}\left(1-e^x\sum_{k=0}^N\frac{(-x)^k}{k!}\right)\\
&=&N!\,\left(\frac{1}{(-x)^{N+1}}-e^x\sum_{j=0}^N\frac{(-x)^{-j-1}}{(N-j)!}\right)
\end{eqnarray*}
Moreover using multiple integration by parts, we can show that:
\begin{eqnarray*}
\int_0^1\!dt\,t^N e^{xt}=\frac{e^x}{x}-\frac{N}{x^2}e^x+\frac{N(N-1)}{x^3}e^x-\ldots-\frac{N!}{(-x)^{N+1}}e^x+\frac{N!}{(-x)^{N+1}}
\end{eqnarray*}
By inspection these two sums are equal and thus we have proven the identity (\ref{identity}).
\bibliography{perturbationtheorypaperv2}

\end{document}